\documentclass[twocolumn]{revtex4-1}
\usepackage{amscd,amssymb,amsmath,latexsym,bm}
\usepackage[mathcal,mathscr]{euscript}
\usepackage{lipsum}
\usepackage{amsfonts}
\usepackage{amsmath}
\usepackage{amssymb}
\usepackage{txfonts}
\usepackage{pxfonts}
\usepackage{graphicx}
\usepackage{movie15}
\usepackage{bm,units,yfonts}
\usepackage{subcaption}
\usepackage[table]{xcolor}
\usepackage{hyperref}
\usepackage{lineno}
\usepackage[nohug]{diagrams}
\diagramstyle[labelstyle=\small]

\newcommand{\CM}{{\mathbb C}}

\newcommand{\NM}{{\mathbb N}}

\newcommand{\RM}{{\mathbb R}}
\newcommand{\SM}{{\mathbb S}}

\newcommand{\ZM}{{\mathbb Z}}

\newcommand{\Aa}{{\mathcal A}}

\begin{document}

\title{Topological Edge Modes by Smart Patterning}

\author{David J. Apigo$^1$, Kai Qian$^2$, Camelia Prodan$^1$ and Emil Prodan$^{2}$}

\affiliation{
$^1$Department of Physics, New Jersey Institute of Technology, Newark, NJ 07102, USA \\
$^2$Department of Physics, Yeshiva University, New York, NY 10016, USA
}

\begin{abstract}
The research in topological materials and meta-materials has reached maturity and is gradually entering the phase of practical applications and devices. However, scaling down experimental demonstrations presents a major challenge. In this work, we study identical coupled mechanical resonators whose collective dynamics are fully determined by the pattern in which they are arranged. We call a pattern topological if boundary resonant modes fully fill all existing spectral gaps whenever the pattern is halved. This is a characteristic of the pattern and is entirely independent of the structure of the resonators and the details of the couplings. The existence of such patterns is proven using $K$-theory and exemplified using a novel experimental platform based on magnetically coupled spinners. Topological meta-materials built on these principles can be easily engineered at any scale, providing a practical platform for applications and devices.
\end{abstract}

 \maketitle
 

\section{Introduction}

Experimental demonstrations of topological effects in classical mechanical systems abound \cite{Prodan2009wq,BergPRE2011vy,Zhang2011,
Kane201339,Chen201413004,KhanikaevNatComm2015,
Deymier2015700,Mousavi2015,Peano2015,Paulose2015153,
Xiao2015920,Paulose20157639,Wang2015,Xiao2015240,
Mao2015,Kariyado2015,Nash201514495,
SusstrunkScience2015,SusstrunkE4767,Deymier2016,
Pal2016,Salerno2016,RocklinPRL2016,PDKPP2017} and the field is rapidly moving towards the next stage, where practical devices and concrete applications ought to emerge. However, maintaining the control over the designs while the systems are scaled down to meet the practical constraints is difficult. In this work, we demonstrate that topological boundary modes can emerge solely from a smart patterning of a meta-material. For certain patterns, without any tuning of the couplings, except for the opening of gaps in the bulk resonant vibrational spectrum, all the bulk spectral gaps of a meta-material become completely filled with topological edge spectrum when the system is halved. This edge spectrum cannot be gapped by any boundary condition or by adiabatic deformation of the meta-material. Due to such minimal tuning, meta-materials designed on these principles may be fabricated at any scale, hence providing a viable pathway towards concrete practical applications.

\vspace{0.2cm}

The goal of our paper is twofold: On one hand, we seek to demonstrate the experimental manifestation of the topological boundary modes in one such smartly patterned meta-material and, on the other hand, we want to explain the theoretical principles behind these unusual predictions. For the first part, we introduce a novel experimental platform based on magnetically coupled spinners. Its hallmark feature is that arbitrarily complex mechanical resonators and couplings can be built by engineering one degree of freedom at a time (see Section~\ref{Sec:CoupledSpinners}). The experimental platform not only enables the realization and characterization of a topological pattern of mechanical resonators, but also helps with the formulation and exemplification of the theoretical concepts, which otherwise may appear quite abstract. Indeed, such topological patterns are necessarily aperiodic, thus the traditional Bloch-Floquet analysis unavailable. The natural theoretical tool to use in such situations is the $K$-theory of $C^\ast$-algebras, as introduced in the pioneering works of \cite{BellissardLNP1986jf} and \cite{KRS-BRevMathPhys2002}. Based on this formalism, we formulate a $K$-theoretic bulk-boundary principle for generically patterned resonators. This principle enables one to resolve the precise conditions in which topological edge spectrum emerges as well as the mechanism behind this phenomenon. Prediction of topological patterns then becomes routine. For simplicity, the present study is restricted to one-dimensional patterns but generalizations to higher dimensions can be easily achieved based on \cite{ProdanSpringer2016} (see {\it e.g.} \cite{ProdanArxiv2018}). 

\vspace{0.2cm}

We believe these theoretical tools would be an extremely useful addition to the materials scientist toolbox and we think that, by combining the somewhat abstract concepts with the concrete experiments, we have finally found a formula to explain the framework to a broad scientific community.

The paper is organized as follows. In Section~\ref{Sec:CoupledSpinners}, the experimental platform based on magnetically coupled spinners is introduced. In particular, we show how to quantitatively map the coupling functions and how to obtain the dynamical matrices that drive the dynamics of the collective resonant modes when the spinners are assembled in arbitrary patterns. This concrete setting is used to explain what topological classification over a fixed pattern means and to give the first hints to why such a program is feasible, despite the fact that the pattern can be arbitrary. In particular, using elementary calculations, we demonstrate that all dynamical matrices take a very specific form, involving only a small set of operators. This leads us in Section~\ref{Sec:AlgApproach} to introduce the algebra of bulk physical algebras, which generates all possible dynamical matrices over a pattern via a canonical representation. For the proposed patterns, this algebra is computed explicitly and shown to be isomorphic to the algebra generated by magnetic translations. Section~\ref{Sec:BulkAnalysis} illustrates the bulk spectra of the proposed patterns and points to the similarity with the Hofstadter butterfly \cite{Hofstadter1976}. We also illustrate the good agreement between numerical and experimental mappings of the spectra. A review of Bellissard's gap labeling procedure \cite{Bel95} is used to rationalize the complexity of the spectra and we show how to compute bulk topological invariants solely from the integrated density of states. It is at this point where the $K$-theory is introduced. Section~\ref{Sec:EdgeStates} is dedicated to the edge analysis. We illustrate the manifestation of the topological edge spectrum through both numerical simulations, quantitative experimental measurements as well as video recordings. Additionally, we explain the $K$-theoretic bulk-boundary mechanism, which culminates with the proof that, indeed, the proposed patterns are topological. A discussion of possible applications concludes the exposition.

\vspace{0.2cm}

Lastly, let us relate our work with the existing literature. To the authors' knowledge, the emergence of topological boundary spectrum in quasi-periodic structures was first pointed out theoretically and observed experimentally in \cite{KLR2012,VZK2013}. Questions about these systems were raised in  \cite{MBB}, namely, were the topological characteristics noted in \cite{KLR2012} a property of one pattern or of an ensemble of patterns? \cite{Pro} provided answers using some of the algebraic methods employed in the present work. By passing from numerical topological invariants to the $K$-theory groups, we answer those questions completely while providing a global chart of all possible topological systems over a quasi-periodic pattern. Additionally, we provide the first experimental observation of such topological edge modes in a quasi-periodic mechanical system. The authors emphasize that this work is part of the vigorous effort of the meta-materials community on the search for topological boundary resonances in aperiodic systems\cite{KLR2012,MBB,KRZ,Pro,HPW,KZ,VZK2013,TGB,TDG,VZL,LBF,DLA,BRS,BLL,AS17,KP2017,BP2017,MNH,LSPZB2018,ZHGW2018,MaoPRL2018}.
This exploration goes well beyond the periodic table of topological insulators and superconductors \cite{SRFL2008,RSFL2010,Kit2009}.

\section{Magnetically Coupled Spinners: A Versatile Experimental Platform}
\label{Sec:CoupledSpinners}

Utilizing a novel experimental platform which can be easily reconfigured and quantitatively characterized enables one to engineer patterns of coupled mechanical resonators with predefined internal structures and couplings, and to study the  spectral properties of virtually any imaginable discrete model. The experimental control over these system is extremely high, resulting in excellent agreement between theory and experiment. One  particular configuration is discussed in detail to exemplify the experimental procedures and is used to demonstrate a topological pattern. The experimental platform will also be used to introduce the program of topological classification over an aperiodic pattern and other more abstract concepts.

\subsection{Coupling and dynamics} 
\begin{figure}
\includegraphics[width=\linewidth]{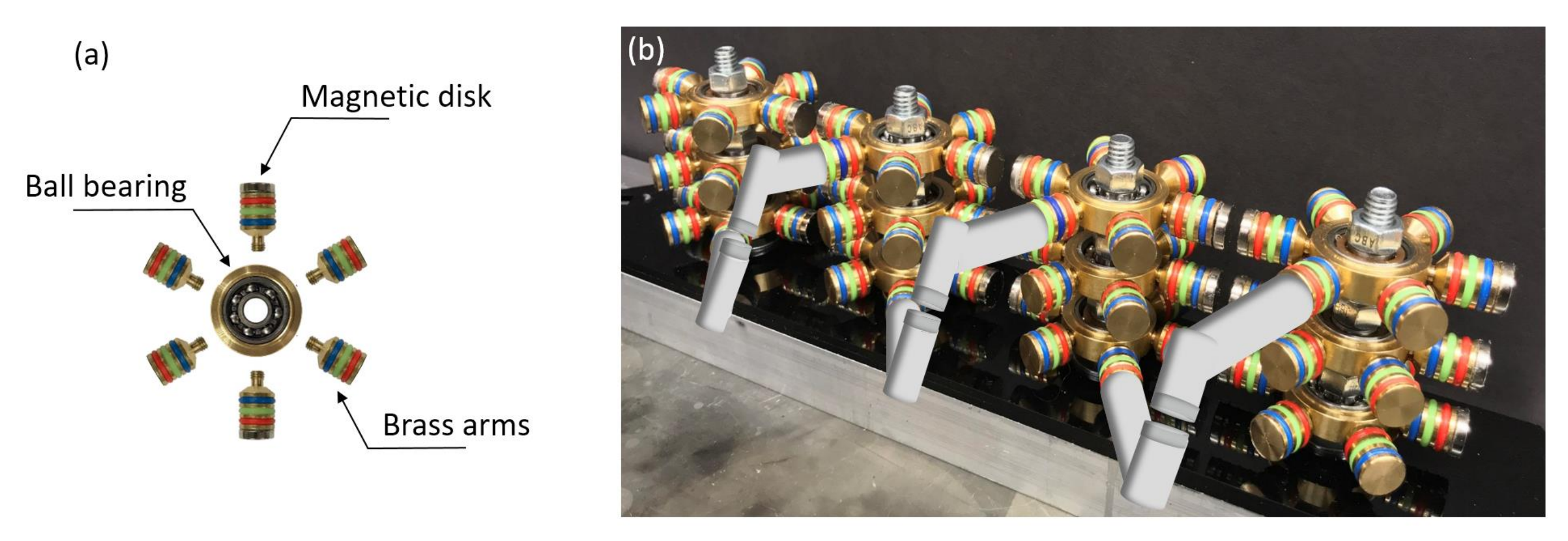}
\caption{\small {\bf Magnetically coupled spinners: A versatile experimental platform.} (a) Example of the basic spinner configuration used in the present work. The arms are detachable such that the spinners can be easily refitted. (b) Exemplification of a relative complex linear configuration of coupled spinners, with coupling in front sideway (shown) and in the back (not shown).}
\label{Fig:Platform}
\end{figure}

A configurable spinner is illustrated in Fig.~\ref{Fig:Platform}(a). It consists of a stainless steel ball-bearing mounted in a threaded brass encapsulation. This enables the spinners to be adorned with a multitude of components. The centers of the spinners are pinned down, resulting in one rotational degree of freedom, $\varphi$. By stacking and coupling such spinners, extremely complex systems can be built one degree of freedom at time (illustrated in Fig.~\ref{Fig:Platform}(b)). With full control over the degrees of freedom and couplings, any quadratic Hamiltonian can be implemented to drive the small oscillations of the coupled spinners. 

\vspace{0.2cm}

\begin{figure}
\includegraphics[width=\linewidth]{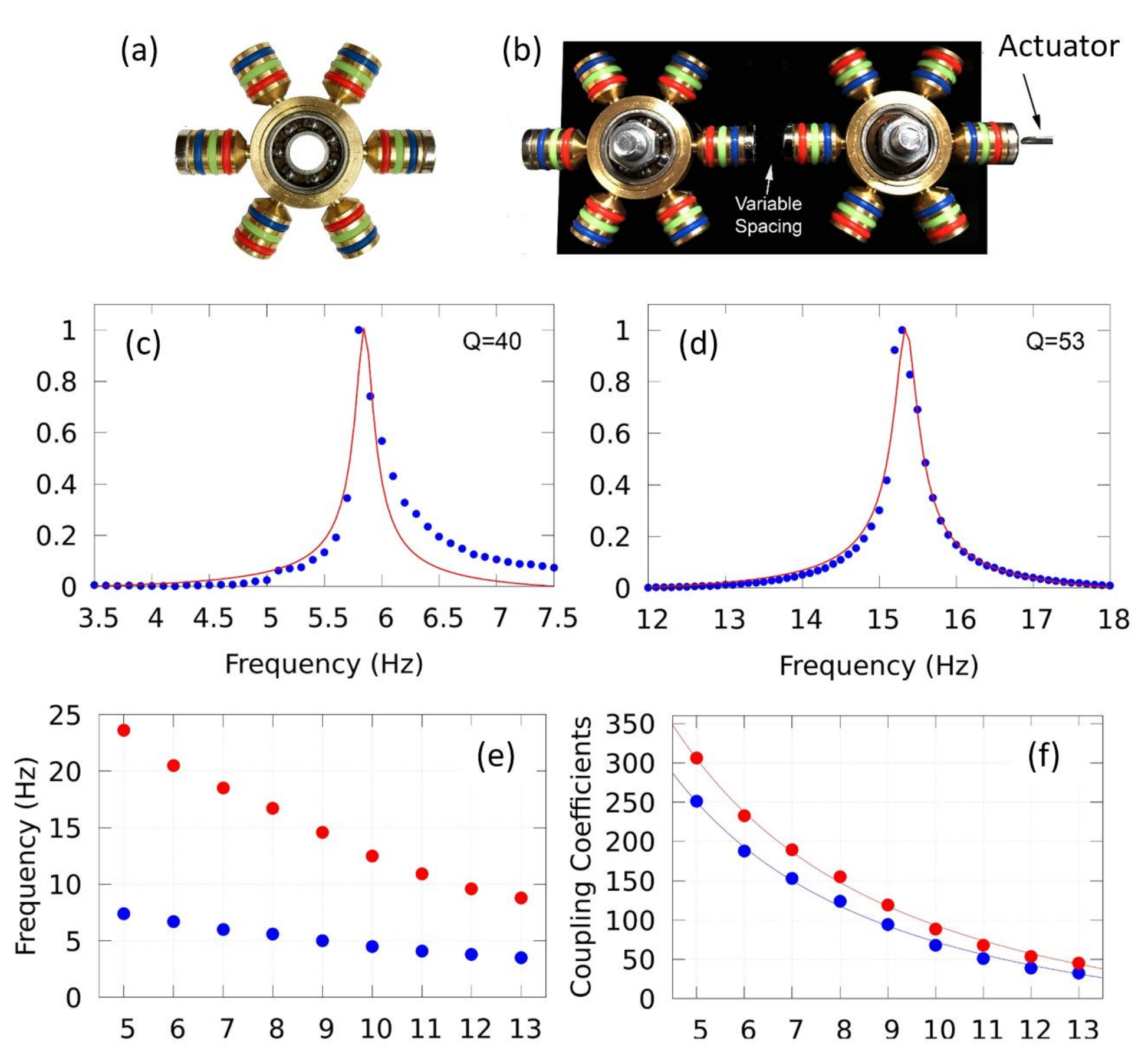}
\caption{\small {\bf Mapping the coupling coefficients.} a) Illustration of a single spinner with the locations of magnets indicated. b) Experimental apparatus for measuring the interaction potential of a dimer. Two spinners are placed on an aluminum track with variable distance, $d$. The system is actuated and response is recorded using accelerometers. c,d) Experimentally measured low/high resonance of a dimer spaced 9 mm apart. The standard fits indicate a quality factors of $Q=40/53$, respectively. e) Map of the high (red) and low (blue) resonant modes as functions of magnet spacing. f) The coupling functions $\alpha$ and $\beta$ as derived from \eqref{Eq:CouplingCoeff}.}
\label{Fig:Coupling}
\end{figure}

The present work features a spinner with six grooved indentations and with heavy brass arms securely fastened in the brass encapsulation. Two of the arms are fitted with neodymium magnetic disks, which provide the couplings between each spinner when arranged in linear patterns. These magnetic couplings can be measured by mapping the resonant modes of a dimer, whose dynamics is governed by the Lagrangian ($I=$ moment of inertia):
 \begin{equation}
L(\varphi_1,\varphi_2,\dot \varphi_1,\dot \varphi_2) = \tfrac{1}{2} I \dot \varphi_1^2 + \tfrac{1}{2} I \dot \varphi_2^2 - V(\varphi_1,\varphi_2).
\end{equation}
In the regime of small oscillations around the equilibrium configuration $\varphi_1=\varphi_2=0$, the potential can be approximated quadratically:
\begin{equation}\label{Eq:QuadraticV}
V(\varphi_1,\varphi_2) = \tfrac{1}{2} \alpha (\varphi_1^2 + \varphi_2^2) + \beta \varphi_1 \varphi_2,
\end{equation}
and the pair of the two resonant modes can be computed explicitly:
\begin{equation}\label{Eq:DimerFreq}
f_{\pm} = \sqrt{\frac{\alpha \pm \beta}{4 \pi^2 I}}.
\end{equation}
The measured resonant frequencies are reported in Fig.~\ref{Fig:Coupling} as functions of distance $d$ between the magnets. Eqs.~\ref{Eq:DimerFreq} can be inverted:
\begin{equation}\label{Eq:CouplingCoeff}
\alpha = 2 \pi^2 I \, \big (f_+^2 + f_-^2 \big ), \quad \beta = 2 \pi^2 I \, \big (f_+^2 - f_-^2 \big ),
\end{equation}
which, together with the experimental data, enable us to determine the functional dependencies $\alpha(d)$ and $\beta(d)$ of the coupling coefficients. The details are provided in Fig.~\ref{Fig:Coupling} and note that units of $2 \pi^2 I$ are used henceforth for the coupling functions.

\vspace{0.2cm}

 When the centers of the spinners are pinned in a one-dimensional pattern $\omega= \{x_n\}_{n\in \mathbb Z}$, the Lagrangian of the system becomes:
\begin{equation}\label{Eq:Lagrangian1}
L = \sum_{n \in \mathbb Z} \bigg [ \tfrac{1}{2} I \dot \varphi_n^2 - \big (\alpha(d_n-1) + \alpha(d_n) \big ) \varphi_n^2 - \beta(d_n) \varphi_n \varphi_{n+1} \bigg ],
\end{equation}
where $d_n = x_{n+1} - x_n - D$ with $D$ the diameter of a spinner. The equations of motion read:
\begin{equation}
-I \ddot \varphi_n = \big (\alpha(d_{n-1}) + \alpha(d_n) \big ) \varphi_n + \beta(d_{n-1}) \varphi_{n-1} + \beta(d_n) \varphi_{n+1}.
\end{equation}
The degrees of freedom can be encoded in the column vector:
\begin{equation}
|\varphi \rangle = ( \ldots, \varphi_{-1},\varphi_0,\varphi_1,\ldots )^T,
\end{equation}
and let us denote by $|n\rangle $ the column vector with 1 at position $n$ and zero in the rest. Then: $|\varphi\rangle = \sum_n \varphi_n |n \rangle$ and, with the ansatz $|\varphi(t) \rangle ={\rm Re}\big [e^{i 2 \pi f t} |\psi\rangle \big ]$ and the units from Fig.~\ref{Fig:Coupling}, the system of equations of motion becomes $f^2 |\psi \rangle = H |\psi \rangle$ with:
\begin{align}\label{Eq:Ham1}
H = \sum_{n \in \mathbb Z} & \bigg[  \big (\alpha(d_{n-1}) + \alpha(d_n) \big ) |n\rangle \langle n | \\ \nonumber
& \quad + \beta(d_{n-1}) |n\rangle \langle n-1| + \beta(d_n) |n \rangle \langle n+1| \bigg ].
\end{align}
It is now a classical eigen-system for Hamiltonian, $H$, in the Hilbert space $\ell^2(\mathbb Z)$. 

\vspace{0.2cm}

We present this analysis in detail because it serves as a model for generically patterned resonators. For example, it can be implemented for other spinner configurations, even for complex ones that include stacking and couplings beyond nearest-neighbors. Throughout this manuscript, especially when discussing continuous deformations of the systems, it is extremely helpful to have a physical realization in mind.

\vspace{0.2cm}

\subsection{Aperiodic yet fully classifiable} 
To resolve the bulk-boundary correspondence principle for these systems, one must deal with the classification of gapped bulk Hamiltonians over the pattern, $\omega$. To properly define the pattern, recall that the spinners can be easily reconfigured, hence the Hamiltonian \eqref{Eq:Ham1} is only one of many that can be implemented over $\omega$. When $N$ spinners are stacked at each point of the pattern, the Hilbert space becomes $\mathbb C^N \otimes \ell^2(\mathbb Z)$, with elementary vectors of the form $\xi \otimes |n\rangle$, and the most general Hamiltonian driving the small oscillations of the coupled spinners takes the form:
\begin{equation}
H_\omega = \sum_{n,n'} w_{n,n'}(\omega) \otimes |n\rangle \langle n' |, 
\end{equation}
where $w_{n,n'} \in M_{N}(\mathbb C)$ with $w_{n',n} = w_{n,n'}^\dagger$. Throughout, $M_N(\mathbb C)$ denotes the space of $N\times N$ matrices with complex entries. The above expression allows couplings beyond the first nearest neighbors and allows for the coupling $N\times N$ matrices to depend on arbitrarily many geometrical data from the pattern $\omega$. The coupling coefficients can be changed continuously (e.g. by modifying the strength of the magnets). Furthermore, by stacking a large number of spinners on top of each other at each point of the pattern, one can smoothly activate or de-activate internal degrees of freedom, changing the dimension $N$. These will be the allowed continuous deformations of our  physical systems. It is useful to view a gapped Hamiltonian as a pair $(H_\omega,G)$, where $G$ is a connected component of the resolvent set $\mathbb R \setminus {\rm Spec}(H_\omega)$. Two gapped Hamiltonians $(H_\omega,G)$ and $(H'_\omega,G')$ are said to be in the same topological class if a continuous gapped deformation connecting the two Hamiltonians exists. The topological classification of the gapped Hamiltonians consists of enumerating these topological classes as well as spelling out at least one representative for each class.

\vspace{0.2cm}

Considering generic aperiodic patterns, the topological classification may appear a daunting task. To understand why this classification is achievable, several key observations are in place:
 
\begin{enumerate}

\item The pattern $\omega$ needs to be treated as an ordinary variable. It takes values in the space of point patterns, a space that can be characterized and topologized using procedures that by now are quite standard \cite{SadunBook}. Existence of a topology is important in defining what a continuous deformation of a point pattern is.

\item Since the spinners are identical copies of a basic design, once the internal structure of the basic spinner is set, the functional dependencies of the couplings on $\omega$ are fixed. More precisely, if the pattern is changed to $\omega'$, we will use the same functions $w_{n,n'}$ but evaluate them at $\omega'$. 

\item Per previous observation, a better terminology for the coupling coefficients would be coupling functions. This is a useful concept because $\omega$, as a point in the space of patterns, contains all the geometric information of the pattern. In general, the coupling coefficients depend on many geometric details of the pattern. If the concept of coupling functions is adopted, then those complicated dependencies can be written concisely as $w_{n,n'}(\omega)$.

\item The coupling functions are assumed to be continuous of $\omega$. We will also consider cases where $w_{n,n'}(\omega)$ becomes negligible for $n$ and $n'$ far apart. Both assumptions are usually met in practice.

\end{enumerate}
The particular Hamiltonian \eqref{Eq:Ham1} reflects all these principles, through the fact that $\alpha(d)$ and $\beta(d)$ have been measured once and then properly evaluated and applied to the arbitrary pattern, $\omega$. The analysis and the measurements that led to \eqref{Eq:Ham1} can be repeated for more complex spinner structures and couplings, and the principles will emerge again. While they seem obvious, these observations bring a unique perspective, which is key to solving the topological classification. 

\vspace{0.2cm} 

Gearing towards that solution, note the natural action of the $\mathbb Z$ group on the space of one-dimensional patterns:
\begin{equation}
\mathbb Z \ni a \rightarrow \tau_a \omega = \tau_a \{x_n\}_{n \in \mathbb Z} = \{x_{n+a}-x_a\}_{n \in \mathbb Z}.
\end{equation}
We will always fix the point labeled by 0 at the origin of the real axis and set the labels to be consistent with the ordering $\ldots <x_{-1}<x_0=0<x_1< \ldots$. This implicitly assumes that two points are never on top of each other. Then $\tau_a$ can be identified with the rigid translation of the pattern that brings point $x_a$ at the origin. Galilean symmetry requires that:
\begin{equation}
w_{n-a,n'-a}(\tau_a \omega) = w_{n,n'}(\omega)  \Rightarrow  w_{n,n'}(\omega) = w_{0,n'-n}(\tau_n\omega). 
\end{equation}
Dropping one redundant index and using $q=n'-n$, as well as the shift operator:
\begin{equation}
S |n \rangle = | n-1 \rangle, \quad S^\dagger |n\rangle = | n+1 \rangle, \quad S S^\dagger = S^\dagger S = I,
\end{equation} 
the generic Hamiltonian takes the form:
\begin{equation}\label{Eq:Ham2}
H_\omega = \sum_q \sum_n w_q(\tau_n \omega) \otimes |n \rangle \langle n |\,  S^q.
\end{equation}  
This expression already reveals a very particular structure. Also, in order to reproduce $H_\omega$, one only needs to evaluate the coupling functions on a small subset of the space of patterns, namely:
\begin{equation}
\Xi = \overline{\big \{ \tau_n \omega, \  n \in \mathbb Z \big \}},
\end{equation}
where the over-line indicates the topological closure of the otherwise discrete set of translated patterns. In the professional literature \cite{SadunBook}, the tuple $(\Xi,\tau)$, which is a bona-fide topological dynamical system, is called the discrete hull of the pattern. A system is called homogeneous if the orbit $\{\tau_n \omega'\}_{n\in \ZM}$ is dense in $\Xi$ for any pattern $\omega' \in \Xi$. We will be dealing exclusively with homogeneous patterns.

\vspace{0.2cm}

The main conclusion is that every Hamiltonian over the pattern $\omega$ can be generated using just the shift operator $S$ and diagonal operators of the form $\sum_n f(\tau_n \omega) |n \rangle \langle n |$ with $f$ a continuous function over $\Xi$. In many instances, the algebra generated by these operators is very simple and connects to other well-known and well-studied algebras. Ultimately, completing the topological classification program over a point pattern is conditioned by the ability to resolve the topological set $\Xi$ and the action of $\ZM$ on it.

\vspace{0.2cm}

\begin{figure}
\includegraphics[width=\linewidth]{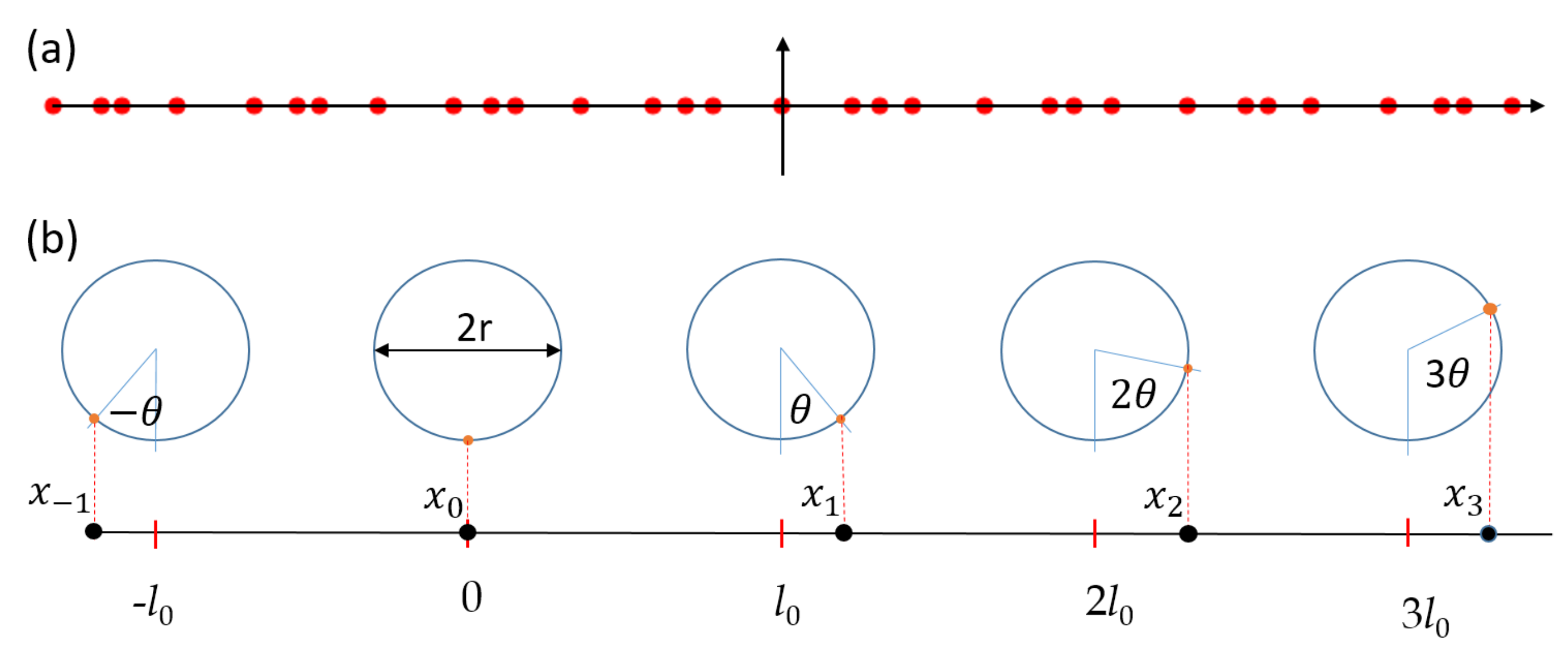}
\caption{\small {\bf Example of a pattern with discrete hull equivalent to $\mathbb S^1$.} a) Display of a finite number of points of a pattern generated by the algorithm $x_n=n\, l_0 +r\, \sin(n \theta)$, with the particular values $l_0=1$, $r=0.4$ and $\theta = \frac{2 \pi}{\sqrt{15}}$. b) A geometric algorithm to generate the same pattern, revealing that $\Xi \simeq \mathbb S^1$.}
\label{Fig:Pattern1}
\end{figure}

\subsection{Generating patterns with prescribed hull}
\begin{figure}
\includegraphics[width=\linewidth]{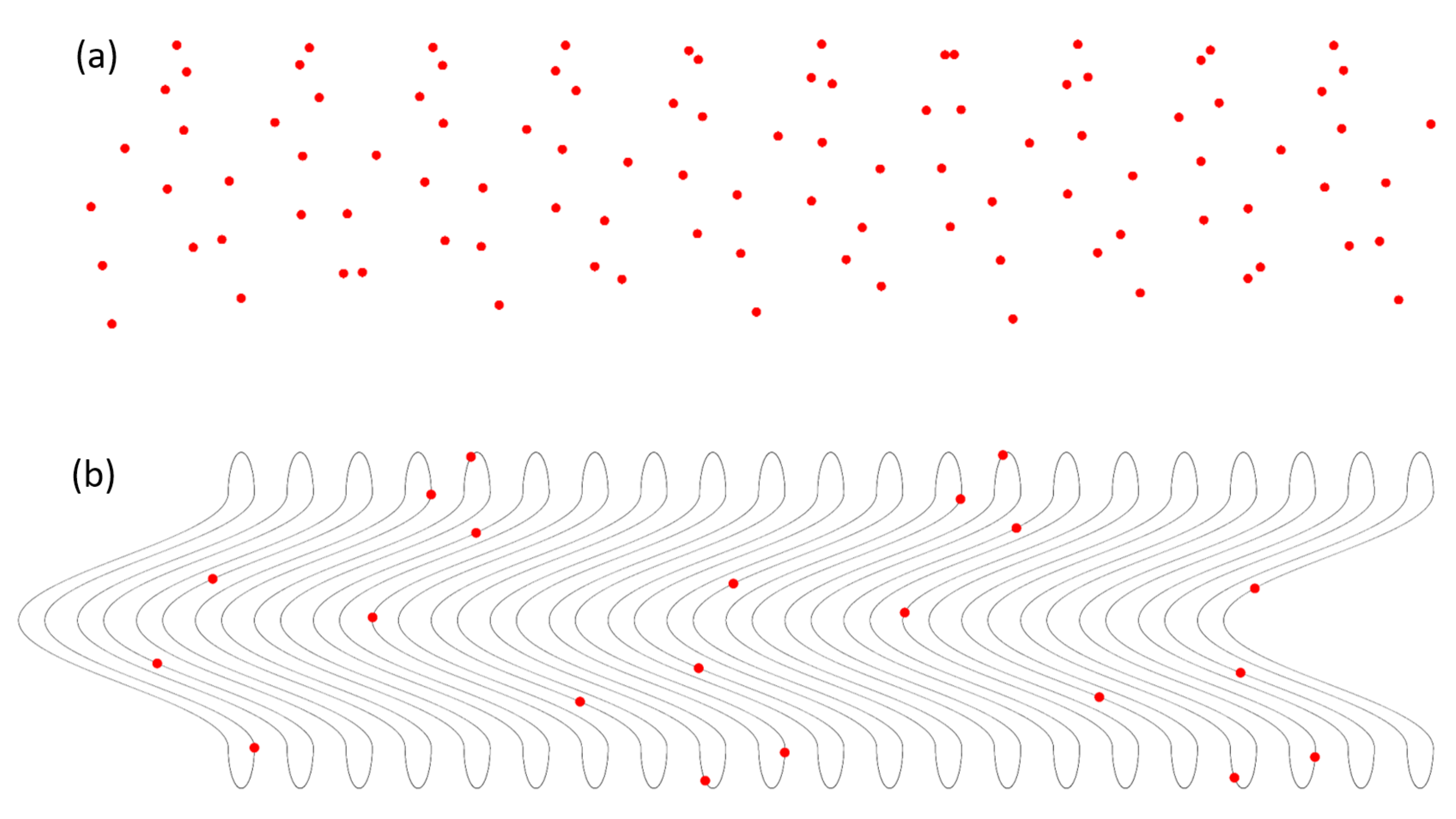}
\caption{\small {\bf Additional example of a pattern with discrete hull equivalent to $\mathbb S^1$.} a) Display of a finite number of points. b) The geometric algorithm used to generate the pattern consists of translation by $\theta$ along the loop followed by a horizontal translation by one unit.}
\label{Fig:Pattern2}
\end{figure}

Here, we generate a class of patterns for which $\Xi$ is topologically equivalent with the circle $\mathbb S^1$. The simplest pattern is illustrated in Fig.~\ref{Fig:Pattern1}(a) and it has the analytic expression:
\begin{equation}\label{Eq:Pattern1}
x_n = n \, l_0 + r\, \sin(n \theta), \quad r < \frac{l_0}{2}, \quad n \in \mathbb Z.
\end{equation} 
The geometric algorithm explained in Fig.~\ref{Fig:Pattern1}(b) can be used to formally derive that $\Xi \simeq \mathbb S^1$. Consider a rigid translation $\tau_a \omega$ of the pattern, such that the old $x_a$ now sits at the origin of the real axis. Associated this $x_a$ there is a point on the circle and it is evident that knowing where this point is located enables us to reproduce the entire translated pattern $\tau_a \omega$. Applying the geometric algorithm described in Fig.~\ref{Fig:Pattern1}(b) (starting from angle $a \theta$ instead of 0) establishes a one-to-one relationship between the translated patterns $\tau_a \omega$ and the angles $a \theta$, $a \in \mathbb Z$. For $\theta$ irrational (in units of $2 \pi$), these points densely fill the circle. It is important to note that $\Xi$ is just a topological space and it has no geometry. From topological point of view, any closed loop is also a circle, hence more complex patterns can be generated by the same algorithm but using a deformed circle. Such a pattern is illustrated in Fig.~\ref{Fig:Pattern2}(a). As one can see, although the algorithm is simple, the resulting patterns can be extremely complex and irregular-looking. Using the same arguments, one can quickly see that $\Xi$ is just the closed loop traced in Fig.~\ref{Fig:Pattern2}(b), hence $\Xi \simeq \mathbb S^1$.  

\vspace{0.2cm}

Based on the above remarks, one can see that we can actually allow not only continuous deformations of the resonators but also of the patterns themselves, provided the topology of the hull remains unchanged. This will be assumed from now on and will be covered by our topological classification.

\section{The $C^\ast$-Algebraic approach}
\label{Sec:AlgApproach}

It is instructive to inspect the commutation relations between the basic operators, which for $N=1$ are:
\begin{align}\label{Eq:CommRel}
& \bigg ( \sum_n f(\tau_n \omega) |n\rangle \langle n| \bigg) \, S \\ \nonumber
& \quad = S \, \bigg ( \sum_n f(\tau_{n} \omega) |n-1\rangle \langle n-1| \bigg) \\ \nonumber
& \quad = S \, \bigg ( \sum_n f(\tau_{n+1} \omega) |n\rangle \langle n | \bigg).
\end{align}
This demonstrates that, when conjugating the specific diagonal operators by $S$, $f$ was effectively replaced by $f\circ \tau_1$. This observation enables us to define an abstract algebra which generates all Hamiltonians $H_\omega$ for all $\omega \in \Xi$.

\subsection{The algebra of bulk physical observables.} We will discuss three important aspects even though some might appear technical at first sight. First, the definition of the algebra $\mathcal A$ of bulk physical observables. It is the universal $C^\ast$-algebra generated by the algebra $C_N(\Xi)$ of continuous function over $\Xi$ with values in $M_N(\mathbb C)$ and by a unitary operator $u$ ($uu^\ast=u^\ast u =1$), satisfying the commutation relations which steam directly from \eqref{Eq:CommRel}:
\begin{equation}\label{Eq:AlgCommRel}
f u = u(f \circ \tau_1), \quad \forall \ f \in C_N (\Xi).
\end{equation}
A generic element from this algebra takes the form $a = \sum_q a_q u^q$, where all coefficients $a_q$ are from $C_N(\Xi)$. The canonical representation on $\mathbb C^N \otimes \ell^2(\mathbb Z)$ is provided by:
\begin{equation}
C_N(\Xi) \ni f \rightarrow \pi_\omega(f) = \sum_n f(\tau_n \omega) \otimes |n \rangle \langle n |
\end{equation}
and $u \rightarrow S$, which we already verified in \eqref{Eq:CommRel} to respect the commutation relations of $\mathcal A$. One can see explicitly that $\pi_\omega (h)$, $h= \sum_q w_q u^q$, generates \eqref{Eq:Ham2}. The algebra $\mathcal A$ generates not only the Hamiltonians but all covariant physical observables over the patterns from $\Xi$, that is, the families of operators $\{A_\omega\}_{\omega \in \Xi}$ with the property:
\begin{equation}
S^{-n} A_\omega S^n= A_{\tau_n \omega}, \quad \forall \ \omega \in \Xi.
\end{equation}

\vspace{0.2cm}

Secondly, since one of the main themes is the classification under continuous deformations, we need to introduce a norm for $\mathcal A$ in order to make precise what the latter means. The canonical norm on $\mathcal A$ is:
\begin{equation}\label{Eq:Norm}
\|a\|= \sup_{\omega \in \Xi} \| \pi_\omega(a)\|,
\end{equation}
where on the right is the ordinary operator norm. When completed under \eqref{Eq:Norm}, $\mathcal A$ becomes a separable $C^\ast$-algebra, which for our program is extremely important because these algebras have well-defined topological $K$-theories and their $K$-groups are always countable. In other words, we are assured that we have a sensible and useful topological classification.

\vspace{0.2cm}

Thirdly, there is an important relationship between the spectrum of an element $h \in \mathcal A$ and the spectra of operators $\pi_\omega(a)$ that stems from it. Recall that the resolvent set of $a$ is:
\begin{equation}
{\rm Res}(a) = \{ \lambda \in \mathbb C \ | \ \lambda - a \ \mbox{is invertible in} \ \mathcal A\}.
\end{equation}
The spectrum of $a$ is then ${\rm Spec}(a) = \mathbb C \setminus {\rm Res}(a)$, a definition that actually makes sense for an arbitrary algebra. In general, we have the isomorphism:
\begin{equation}
\mathcal A \simeq \bigoplus_{\omega \in \Xi} \pi_\omega(\mathcal A) \quad \Rightarrow \quad {\rm Spec}(h) = \bigcup_{\omega \in \Xi} {\rm Spec} (H_\omega ).
\end{equation}
However, for a homogeneous system, ${\rm Spec}(H_\omega )$ is independent of $\omega$ and, as such:
\begin{equation}
{\rm Spec}(h)={\rm Spec}(H_\omega) \quad \forall \, \omega \in \Xi,
\end{equation}
a conclusion which will play an important role in our final discussion.

\vspace{0.2cm}

\subsection{Explicit computations.} If the pattern $\omega$ is periodic, then $\Xi$ is a point and the algebra $\mathcal A$ is generated by the shift operator $S$. Hence, it is commutative and we are dealing with the ordinary band theory. 

\vspace{0.2cm}

If $\omega$ is a disordered lattice, {\it i.e.} small random displacements drawn from the interval $[-r,r]$ of otherwise equally spaced points, then $\Xi$ is the Hilbert cube $[-r,r]^{\mathbb Z}$. This space has trivial topology since it is contractible to a point and the $K$-theory of the resulting observable algebra is the same as the $K$-theory of the periodic lattice \cite{ProdanSpringer2016}.

\vspace{0.2cm}

The simplest example with a non-trivial topology is when $\Xi$ is equivalent to the circle and $\tau_1$ is the translation by a fixed $\theta$, as in the examples from Figs.~\ref{Fig:Pattern1} and \ref{Fig:Pattern2}. For $N=1$, we know from the ordinary Fourier analysis that the algebra $C(\Xi)$ is generated by one function, $v(s)=e^{is}$, where $s$ is the coordinate along the $\Xi$, assumed to be a closed loop of length $2 \pi$. The commutation relations can be computed \eqref{Eq:AlgCommRel} explicitly:
\begin{equation}\label{Eq:ExCommRel}
v \, u = u\, (v \circ \tau_1) = e^{i \theta} u \, v,
\end{equation}
because $(v\circ \tau_1)(s)=e^{i(s+\theta)} = e^{i\theta} v(s)$. The conclusion is that $\mathcal A$ is the non-commutative 2-torus. This algebra is the same as the one generated by the magnetic translations, from where one draws the Hamiltonians for electrons hopping on a lattice in a perpendicular uniform magnetic field. The latter is the setting where the Integer Quantum Hall Effect (IQHE) is observed, which is the prototypical topological system from class A in 2-dimensions. As we shall see, there are extremely close spectral and topological similarities between the those systems and the ones studied in this work. 

\vspace{0.2cm}

One important point of these exercises was to convey that if $(\Xi,\tau)$ can be resolved and is simple enough, then the algebra $\mathcal A$ can be computed explicitly. In many cases, it can be connected with already well-studied algebras. In these cases, the classification of the gapped Hamiltonians can be fully carried out and the bulk-boundary principle can be formulated very precisely, as we shall see next.

\section{Deciphering the Bulk}
\label{Sec:BulkAnalysis}

\begin{figure*}
\includegraphics[width=0.9\linewidth]{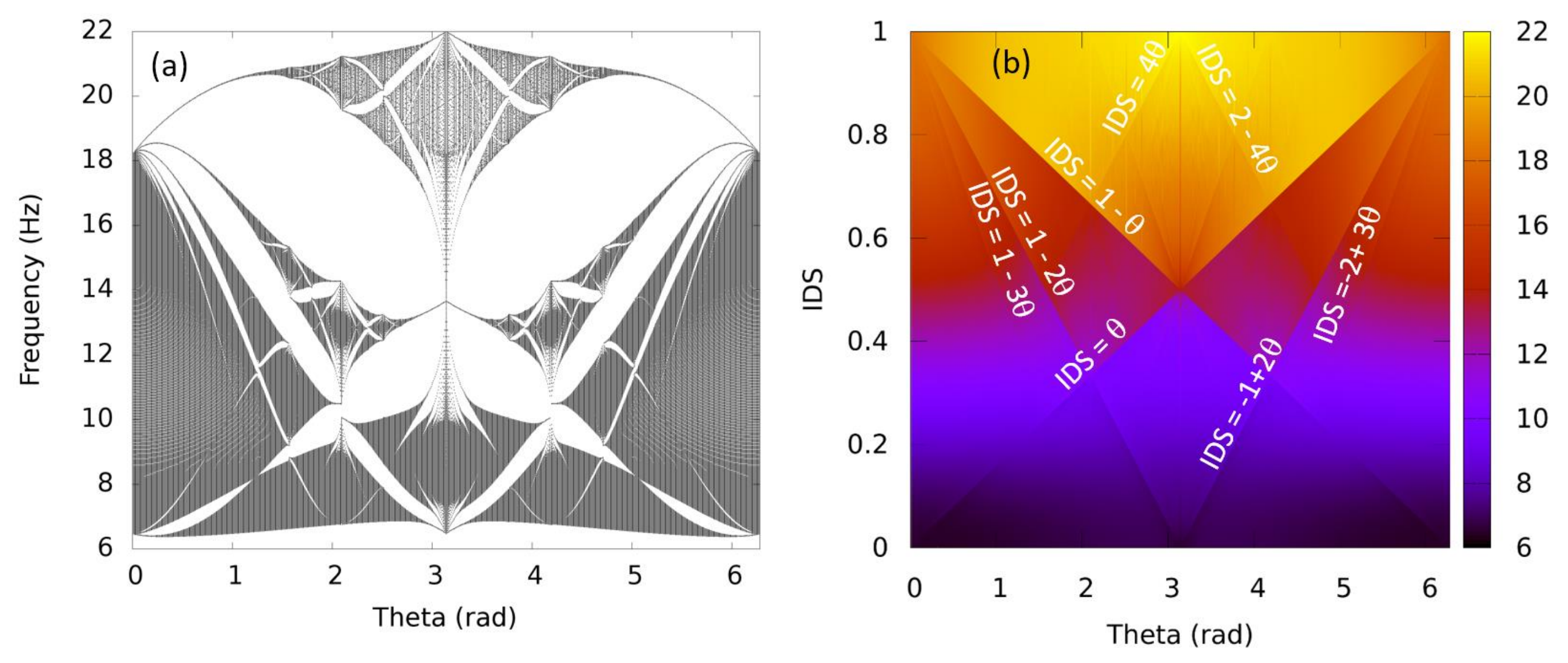}
\caption{\small {\bf Bulk spectral characteristics of the Hamiltonian \eqref{Eq:Ham1} over the pattern from Fig.~\ref{Fig:Pattern1}.} (a) Bulk spectrum as function of parameter $\theta$. The parameters $l_0$ and $r$ have been fixed to the experimental values (see Fig.~\ref{Fig:BulkExpVsTh}). (b) Integrated density of states \eqref{Eq:IDS} rendered as function of $\theta$ and frequency. The frequency axis is out of the plane.}
\label{Fig:BulkSpecIDS}
\end{figure*}

\subsection{Facts and observations.} The resonant frequency spectrum for Hamiltonian \eqref{Eq:Ham1} over the pattern \eqref{Eq:Pattern1} is shown in Fig.~\ref{Fig:BulkSpecIDS}(a) as function of $\theta$. The calculation was performed on a finite pattern of length $L=840$ with periodic boundary condition for all commensurate values $\theta_n = \frac{2 n \pi}{L}$ (note that the spectrum is known to be continuous of $\theta$, hence the use of rational values is not an issue here). The empirical couplings $\alpha$ and $\beta$ have been used in these calculations. The similarity between this spectrum and the Hofstadter spectrum \cite{Hofstadter1976} is remarkable. The main characteristics of the spectrum is the fractal network of spectral gaps. Despite its complexity, the spectral gaps can be labeled uniquely by just two integer numbers \cite{Bel95}. A practical way to achieve this labeling is to compute the integrated density of states (IDS), defined as:
\begin{equation}\label{Eq:IDS}
IDS(f)=\left . \frac{\# \ \mbox{resonant frequencies below} \ f}{\mbox{ Length}\ L} \ \right |_{L \rightarrow \infty}.
\end{equation}
A graphic representation of IDS as function of $f$ and $\theta$ is reported in Fig.~\ref{Fig:BulkSpecIDS}(b), as derived from the data reported in Fig.~\ref{Fig:BulkSpecIDS}(a). In this rendering, the sharp changes in color are associated with the spectral gaps and the value of IDS inside the spectral gaps are all characterized by straight lines:
\begin{equation}\label{Eq:IDSValues}
IDS(\theta) = n + m \, \theta, \quad n,m\in \mathbb Z.
\end{equation} 
Another key observation is that, since $N=1$, the $IDS$ is always bound to the interval $[0,1]$. As we shall see later, the index $m$ in \eqref{Eq:IDSValues} is the topological number which dictates the presence or absence of edge modes. Examining \eqref{Eq:IDSValues}, one sees that there are only two instances where $m=0$; when the states are fully depopulated ($IDS=0$) or fully populated ($IDS=1$). We can now anticipate the main finding of our work; every gap seen in Fig.~\ref{Fig:BulkSpecIDS}(a) is topological in the sense that $m \neq 0$ and this implies the emergence of topological edge states. This is a statement which applies to any Hamiltonian \eqref{Eq:Ham2} with $N=1$. Thus, it is an intrinsic characteristic of the pattern. By all measures, the pattern can be called topological. 

\vspace{0.2cm}

\begin{figure*}
\includegraphics[width=0.9\linewidth]{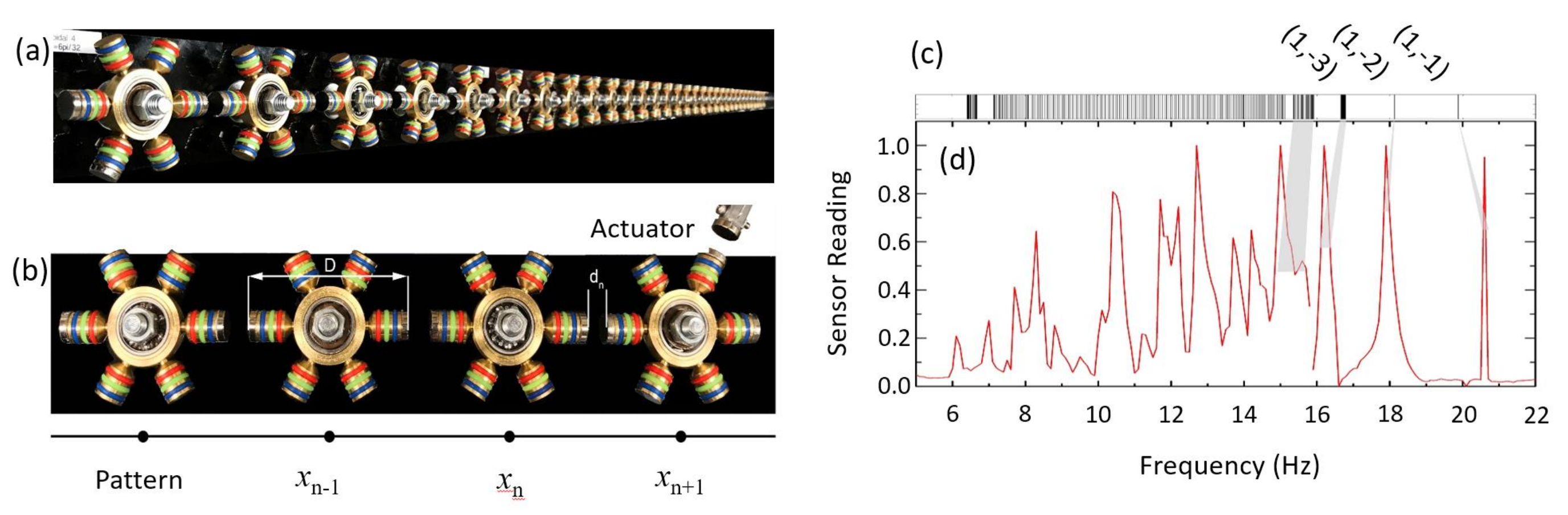}
\caption{\small {\bf Experimental bulk spectral characteristics.} (a) A system of 32 spinners arranged in pattern \eqref{Eq:Pattern1}. (b) Notations and experimental values: $\theta = \frac{6\pi}{32}$, $D=66$~mm, $l_0=76$~mm, $r=2$~mm. (c) Theoretically computed bulk spectrum for $\theta = \frac{2\pi}{\sqrt{117}}=\frac{6\pi}{32} + \mathcal O(10^{-3})$, together with the gap labels for the upper gaps, as extracted from Fig.~\ref{Fig:BulkSpecIDS}. (d) Experimental reading from an accelerometer placed in the bulk of the system. The correspondence between theory and experiment is shown by the shaded regions.}
\label{Fig:BulkExpVsTh}
\end{figure*}

The bulk spectrum has been mapped experimentally for select values of $\theta$. The setup is shown in Figs.~\ref{Fig:BulkExpVsTh}(a,b). Throughout, the units of length are millimeters. To accommodate for the diameter, $D=66$~mm, of the spinners, their centers have been arranged according to the algorithm $x_n = 76n + 2 \, \sin(n \theta)$, $\theta=\frac{6\pi}{32}$, leading to a distance between the magnets (see Eq.~\ref{Eq:Lagrangian1}): 
\begin{equation}\label{Eq:DnAlgorithm}
d_n=10 +2 \sin\big((n+1)\theta\big ) - 2 \sin\big ( n \theta \big ), \quad n \in \mathbb Z. 
\end{equation}
These are the inputs for Hamiltonian \eqref{Eq:Ham1}. The theoretically computed spectrum and the experimentally measured one are reported in Figs.~\ref{Fig:BulkExpVsTh}(c,d), respectively. The overall quantitative agreement is good, especially for the upper part of the spectrum. In fact, a rigorous correspondence between the two was established (see the guiding shaded regions in Figs.~\ref{Fig:BulkExpVsTh}(c,d)), by matching the experimental and theoretical profiles of the normal modes. The IDS is extremely difficult to map experimentally since the size of the system needs to be appreciable and each of the resonant frequencies need to be resolved. For these reasons, no attempt was made towards an experimental measurement.

\subsection{Gap labeling: The $K$-theoretic account.} One extremely puzzling question is how do the straight lines emerge in Fig.~\ref{Fig:BulkSpecIDS}(b), given that no tuning of the coupling coefficients has been attempted? For example, identical straight lines will show up even if the spinners are arranged in the complicated pattern shown in Fig.~\ref{Fig:Pattern2}. The fundamental principle behind this phenomenon is easily explained by the $K$-theory of the observables algebra, $\mathcal A$ \cite{Bel95}.

\vspace{0.2cm} 

Given an element, $h$, of the $C^\ast$-algebra and a continuous function $\varphi:\mathbb C \rightarrow \mathbb C$, one can define a functional calculus $\varphi(h)$ by approximating $\varphi$ by polynomials and taking the limit with respect to \eqref{Eq:Norm}. This limit exists if and only if $\varphi$ is continuous on the spectrum of $h$ (see \cite{ArvesonBook2002}). The example which will often appear from now on is the gap projection $p_G=\chi_{(-\infty,G]}(h)$, where $h$ is the element of $\mathcal A$ which generates a covariant family of gapped Hamiltonians $(H_\omega,G)$. From now on, $G$ will represent the gap itself and also an arbitrary point from the gap. Also, $\chi$ is the characteristic function of the specified interval. In the expression of $p_G$, $\chi_{(-\infty,G]}$ has a discontinuity at $G$ but, since it occurs outside the spectrum of $h$, $p_G$ is indeed an element of $\mathcal A$, which generates the spectral projections of $H_\omega$'s onto the spectrum below $G$. Consider now the family of functions $\varphi_t(x)=\frac{x-G}{1-t+t|x-G|}$, which interpolates continuously between $\varphi_0(x)=x$ and $\varphi_1(x)={\rm sgn}(x-G)$. Then, $\varphi_t(H_\omega)$ interpolates continuously between $H_\omega$ and ${\rm sgn}(H_\omega) = 1- 2 \chi_{(-\infty,G]}(H_\omega)$. What this suggests is that classifying gap Hamiltonians $(H_\omega,G)$ are the same as classifying the projections $P_\omega(G)=\chi_{(-\infty,G]}(H_\omega)$ or, at the level of algebra $\mathcal A$, the projections $p_G$. Typically, there are many projections in an algebra but, if they are organized in topological equivalence classes, their accountability becomes possible. This is what $K$-theory offers \cite{BellissardLNP1986jf,DavidsonBook,WOBook1993,BlaBook1998,RLLBook2000}.

\vspace{0.1cm}

Given a generic $C^\ast$-algebra $\mathcal A$, the $K_0$ group is defined as the classes $[p]_0$ of projections ({\it i.e.} $p^2=p^\ast=p$) from $M_N(\mathbb C) \otimes \mathcal A$ with $N$ arbitrarily large (hence $M_\infty(\mathbb C)$ is used instead), where two projections belong to the same class iff they can be continuously deformed into each other or if there is $u \in M_\infty(\mathbb C) \otimes \mathcal A$ such that $p'=upu^\ast$ (when $\mathcal A$ is tensored by $M_\infty(\mathbb C)$, the two criterion coincide). Given two projections and their classes, one defines $[p]_0 \oplus [q]_0=\begin{pmatrix} p & 0 \\ 0 & q\end{pmatrix}_0$, which makes $K_0$ into an Abelian semi-group, which then can be completed to a group. This is how $K_0(\mathcal A)$ group is defined. Similarly, two unitary elements from $M_\infty(\mathbb C) \otimes \mathcal A$ are declared to be in the same $K_1$-class if they can be continuously deformed into each other. Given two unitaries and their $K_1$-classes, one defines the binary operation $[u]_1 \odot [u']_1 = [uu']_1$, which transforms $K_1(\Aa)$ into an Abelian group. Since the projections and unitaries are drawn from $M_\infty(\mathbb C) \otimes \mathcal A$, we can simplify and take $N=1$ in the definition of the bulk algebra, because $M_\infty(\mathbb C)$ automatically takes care of the internal degrees of freedom!

\vspace{0.2cm}

We now state a central statement. By definition, the class of a projection is a topological invariant, though not a numerical one. As long as we classify the systems by $K$-theory, which is now well understood within our community to be the physically correct way to do (see next paragraph), the class $[p_G]_0$ inside the $K_0$-group is the most general topological invariant that can be associated to a gap projection. Another key observation is that, if $\mathcal A$ is a separable $C^\ast$-algebra as in our case, then both $K_0(\mathcal A)$ and $K_1(\mathcal A)$ are, at most, countable. In fact, for the non-commutative 2-torus, which is the algebra associated to our patterns, $K_0(\mathcal A) \simeq \mathbb Z^2$, hence it has only two generators, the identity $[1]_0$ and the Rieffel projection $[p_\theta]_0$ \cite{Rieff}. As such, up to homotopies, any projection from $M_\infty(\mathbb C)\otimes \mathcal A$ can be decomposed as: 
\begin{equation}
[p]_0=[1]_0\oplus \ldots \oplus [1]_0 \oplus [p_\theta]_0 \oplus \ldots \oplus [p_\theta],
\end{equation}
or simply as:
\begin{equation}
[p]_0= n \, [1]_0+ m \, [p_\theta]_0, \quad n,m \in \mathbb Z.
\end{equation}
Therefore, we can locate $p$ in $K_0(\mathcal A)$ using just the two integers $n$ and $m$. As long as one classifies by $K$-theory, these integers represent the complete set of topological invariants that can be associated to a projection. It remains to show that they are the same numbers appearing in our previous IDS analysis. For this, note that the $IDS$ values inside the gaps can be also computed as the trace per length of the gap projections:
\begin{equation}
IDS(G) = \lim_{N \rightarrow \infty} \frac{1}{2N}\sum_{n=-N}^N \langle n | P_\omega(G) |n \rangle.
\end{equation}
At the level of algebra $\mathcal A$, the trace per length has a very simple interpretation. Indeed:
\begin{align}
\lim_{N \rightarrow \infty} \frac{1}{2N}\sum_{n=-N}^N \langle n | A_\omega |n \rangle & = \lim_{N \rightarrow \infty} \frac{1}{2N}\sum_{n=-N}^N \langle n | \pi_\omega(a) |n \rangle  \\ \nonumber 
& = \lim_{N \rightarrow \infty} \frac{1}{2N}\sum_{n=-N}^N a_0(\tau_n \omega),
\end{align}
and, by using Birkhoff's ergodic theorem \cite{Bir1931}, we can conclude that:
\begin{equation}\label{Eq:Partial1}
\lim_{N \rightarrow \infty} \frac{1}{2N}\sum_{n=-N}^N \langle n | A_\omega |n \rangle = \int_\Xi {\rm d} \mathbb P(\omega) \, a_0(\omega),
\end{equation}
where $\mathbb P(\omega)$ is the unique translation invariant probability measure on $\mathbb S^1$. The right hand side defines a trace on the algebra $\mathcal A$, which will be denoted by $\mathcal T$, {\it i.e.} a positive linear functional such that $\mathcal T(a a') = \mathcal T(a'a)$ for any $a,a' \in \mathcal A$. This trace can be trivially extended over $M_\infty(\mathbb C) \otimes \mathcal A$ by tensoring with the ordinary trace. Now, consider $p$ and $p'$ from the same $K_0$-class. Then there exists $u$ such that $p' = u p u^\ast$ and consequently $\mathcal T(p) = \mathcal T(p')$. As a consequence, the trace $\mathcal T$ is constant over the topological classes, hence it defines a topological invariant. Given the linearity of the trace \cite{BellissardLNP1986jf,Bel95}:
\begin{align}\label{Eq:IDSRange}
IDS(G) & = \mathcal T([p_G]_0) = \mathcal T(n[1]_0\oplus m [p_\theta]_0) \\ \nonumber 
& = n \mathcal T([1]_0) + m \mathcal T([p_\theta]_0) = n + m \theta,
\end{align}
where for the last equality we used the fundamental result $\mathcal T(p_\theta) = \theta$ \cite{Rieff}.

\vspace{0.2cm}

\begin{figure}
\includegraphics[width=\linewidth]{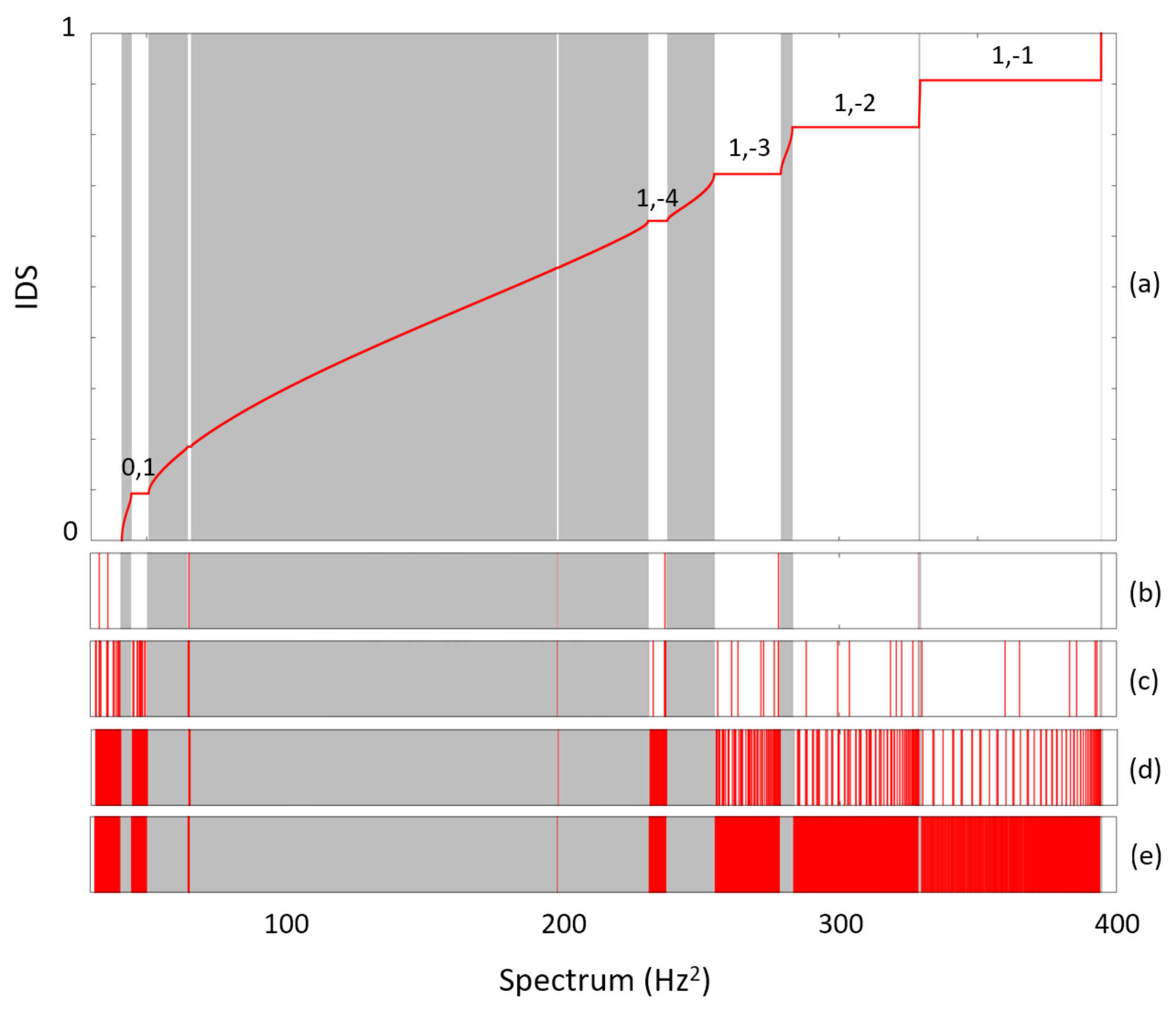}
\caption{\small {\bf Numerical illustration of the topological edge spectrum.} (a) Integrated density of states (IDS) for the pattern \eqref{Eq:DnAlgorithm} with $\theta = \frac{2 \pi}{\sqrt{117}}$ (red curve). Overlaid in gray is the bulk spectrum. The $IDS$ values $n+m \theta$ inside the spectral gaps are indicated by the pairs of integers $(n,m)$. (b-e) Edge spectrum (red marks) of $N_s$ stacked edged systems, with $N_s=1,10,100,1000$, respectively. For convenience, the bulk spectrum is shown in gray.}
\label{Fig:EdgeSpecTh}
\end{figure}

Several observations are in place, with fundamental consequences for experiments. Given any Hilbert space $\mathcal H$, its algebra $\mathbb B(\mathcal H)$ of bounded operators is not separable and its $K$-theory is irrelevant. In fact $K_0(\mathbb B)=0$ for any separable $\mathcal H$ \cite{WOBook1993}), so it is very important that all the Hamiltonians over a pattern can be all drawn from the smaller algebra $\mathcal A$ with a non-trivial $K$-theory. The number of internal degrees of the resonators cannot be fixed in general. For example, in quantum chemistry we use pseudo-potentials and discard the deep electron states, which are chemically inert, and we also get rid of the states in the continuum spectrum. The tight-binding Hamiltonians used to model topological insulators are just effective Hamiltonians where an infinite number of internal states are ``integrated out.'' It is important to acknowledge that, in $K$-theory, states can be added without changing the classification. This is why the $K$-theoretic classification is more physical than any other classification schemes. Now an extremely fine and important point: by definition, the labels $n$ and $m$ cannot be changed as long as $p$ is continuously deformed. This deformation needs to happen inside the algebra, $\mathcal A$, by either deforming the resonators, the way they couple or by deforming the pattern without changing the topology of the discrete hull $\Xi$ (as for the patterns in Figs.~\ref{Fig:Pattern1} and \ref{Fig:Pattern2}). These give the precise experimental conditions in which the predictions based on $K$-theory will hold, something which in the physical literature is completely overlooked, yet they are paramount for the practical applications.

\section{Topological Edge States}
\label{Sec:EdgeStates}

\begin{figure}
\center
  \includegraphics[width=\linewidth]{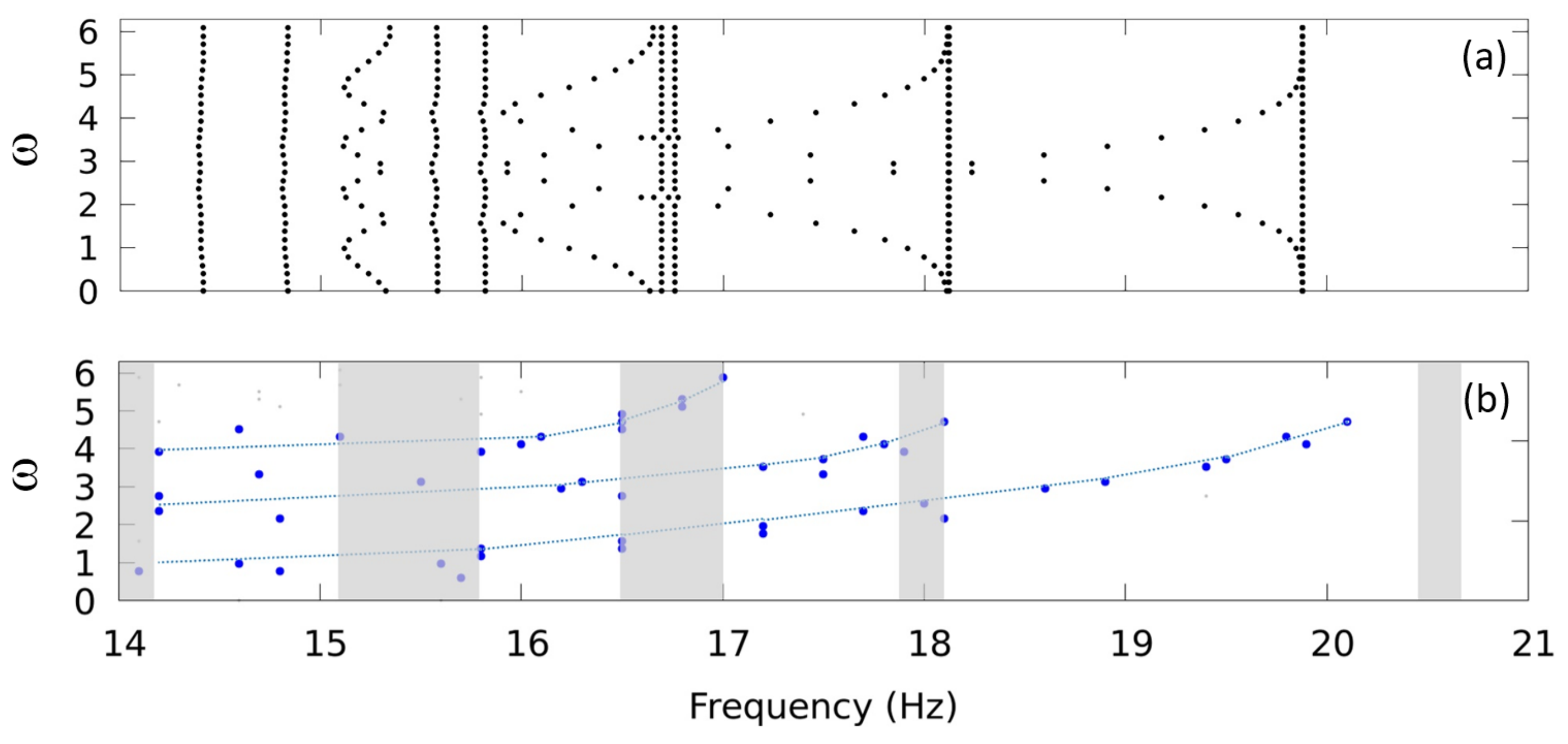}\\
  \caption{\small {\bf Theoretical edge spectrum versus the measured one.} The predicted theoretical edge spectrum is given on top as function of $\omega$. As one can see, chiral bands appear crossing all bulk visible bulk gaps. Below, the experimental data (dots) is shown with the bulk spectrum indicated by vertical grey bars. Dotted lines have been added to help indicate the chiral bands.}
 \label{Fig:EdgeExpVsTh}
\end{figure}

In this section, we remove the degrees of freedom with index $n<0$ and examine the spectral properties of the edged Hamiltonians, like:
\begin{equation}
\widehat H_\omega = \sum_{n,n'\geq 0} w_{n,n'}(\omega) \otimes |n\rangle \langle n' |,
\end{equation}
defined over the Hilbert space $\mathbb C^N \otimes \ell^2(\mathbb N)$. The above Hamiltonians assumes that all spinners with $n<0$ have been jammed. However, the edge can be generated in many different ways. For example, one could remove all the spinners with index $n<0$, in which case some of the coupling coefficients near the edge are altered. In real-world applications, we may never produce and maintain clean edges. Instead, the coupling constants will be drastically affected near the edge by the cutting process or by gradual wear and tear. Hence, it is very important to produce statements that are independent of the boundary conditions. This is another aspect where $K$-theory shows its effectiveness.

\vspace{0.2cm}

 When a bulk system $H_\omega$ is edged to $\widehat H_\omega$, the bulk spectrum remains in place, but additional spectrum can emerged inside the bulk spectral gaps. It is useful to introduce the edge spectrum as:
 \begin{equation}
 {\rm Spec}_e(\widehat H_\omega) = {\rm Spec}(\widehat H_\omega) \setminus {\rm Spec}(H_\omega).
 \end{equation}
 In one and quasi-one dimensional systems, ${\rm Spec}_e(\widehat H_\omega)$ can only contain a finite number of eigenvalues. Interesting things can happen when the systems are stacked as:
 \begin{equation}\label{Eq:Stack}
 \widehat H= \bigoplus_{\omega \in \Omega} \widehat H_\omega.
 \end{equation} 
 The bulk spectrum remains unchanged, but the edge spectrum now consists of:
\begin{equation}\label{Eq:StackedEdgeSpec}
{\rm Spec}_e(\widehat H)=\bigcup_{\omega \in \Xi}{\rm Spec}_e(\widehat H_\omega) = \overline{\bigcup_{a \in \ZM}{\rm Spec}_e(\widehat H_{\tau_a \omega})},
\end{equation}
where in the last equality we used the fact that the orbit of $\omega$ is dense in $\Xi$. The last equality shows that the stacking can be achieved by simply cutting the same chain but at different locations. The edge spectrum is said to be topological if it fills the bulk gap completely:
\begin{equation}
{\rm Spec}_e(\widehat H) \cap G = G,
\end{equation}
 and if it cannot be removed by any adiabatic deformation of the bulk system or by changing the boundary condition. 

\vspace{0.2cm}

 \begin{figure*}
\center
  \includegraphics[width=0.8\linewidth]{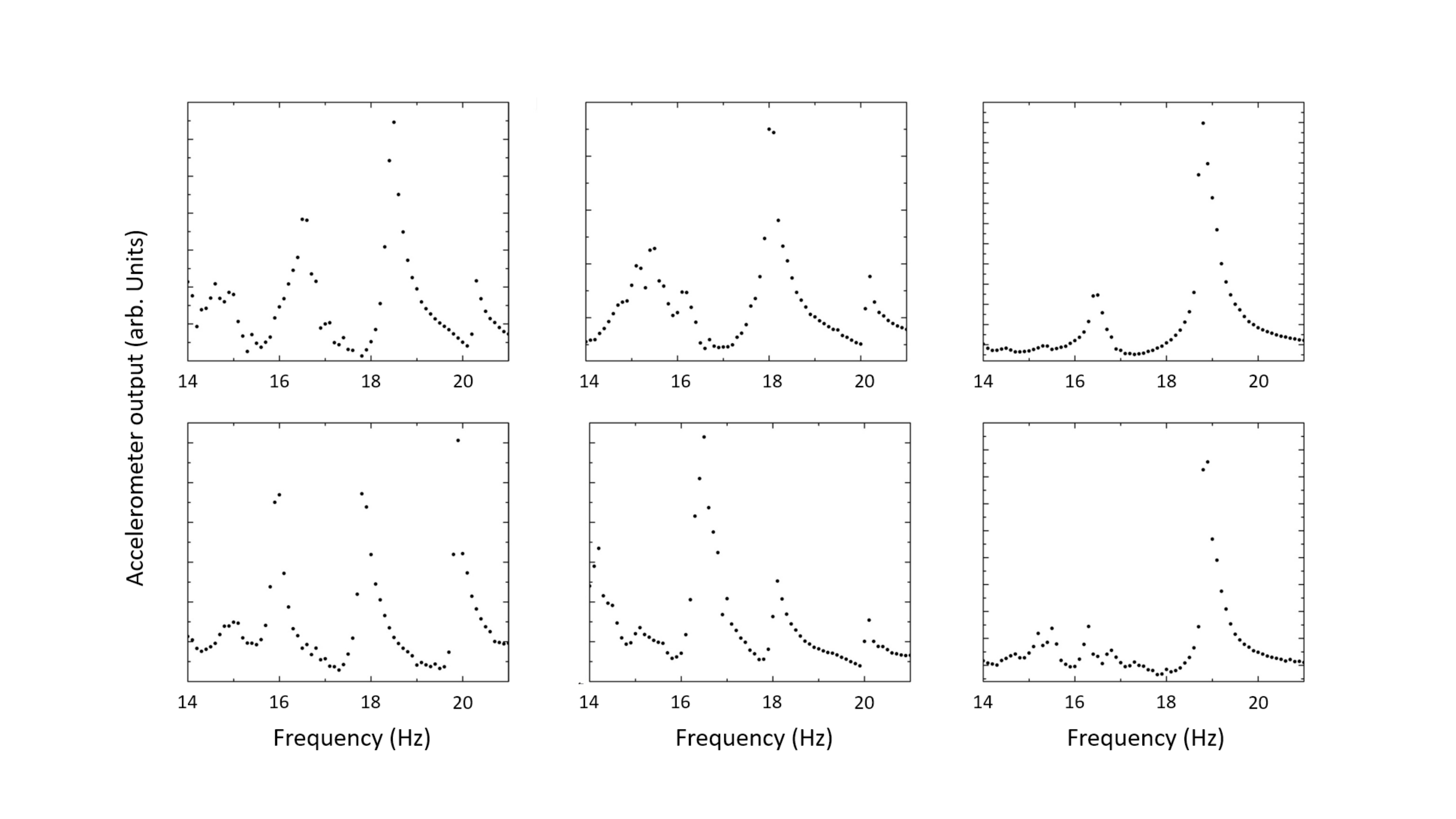}\\
  \caption{\small {\bf Measurements of the edge resonances.} The panels report the readings from accelerometers placed on the second spinner from the edge for randomly rotated configurations of the spinner chain. The edge resonances appear as prominent peaks in these measurements.}
 \label{Fig:EdgeResonance}
\end{figure*}

\subsection{Facts and observations} After removing all the spinners with index $n<0$, the dynamics is described by the Hamiltonian:
\begin{align}\label{Eq:EdgeHam1}
\widehat H_\omega = & \sum_{n \in \mathbb N} \bigg[  \big (\alpha(d_{n-1}) + \alpha(d_n) \big ) |n\rangle \langle n |  \\ \nonumber 
& + \beta(d_{n-1}) |n\rangle \langle n-1| + \beta(d_n) |n \rangle \langle n+1| \bigg ],
\end{align}
with the understanding that $\alpha(d_{-1})=\beta(d_{-1})=0$. In Fig.~\ref{Fig:EdgeSpecTh} we show the theoretically computed spectra of $N_s$ stackings of edged systems:
\begin{equation}
\bigcup_{a=0,\ldots,N_s-1} {\rm Spec}(\widehat H_{\tau_a \omega}), \quad N_s=1,\ 10, \ 100, \ 1000,
\end{equation}
for the pattern \eqref{Eq:DnAlgorithm} with the experimental coupling coefficients. The angle was fixed at $\theta = \frac{2 \pi}{\sqrt{117}} \approx 0.5808$. This particular irrational fraction of $2 \pi$ accepts a good rational approximation $\theta= \frac{6\pi}{32} + \mathcal O(10^{-3})$ and are used in the experiment. In the numerical calculations the exact $\theta$ was used and the calculation was performed on a finite pattern with $L=7669$, chosen based on the 
 rational approximation $\theta=2\pi\frac{709}{7669}+\mathcal O(10^{-7})$. This ensures that all graphical representations in Fig.~\ref{Fig:EdgeSpecTh} are extremely accurate.
 
 \vspace{0.2cm} 
 
Fig.~\ref{Fig:EdgeSpecTh}(a) reports the computed $IDS$ together with the gap labels $(n,m)$ derived from the values of $IDS$ inside the gaps. They are in agreement with the labels seen in Fig.~\ref{Fig:BulkSpecIDS}.  Examining Figs.~\ref{Fig:EdgeSpecTh}(b-e), one can witness how all the spectral gaps of the bulk Hamiltonian are gradually filled with boundary spectrum as more systems are added to the stack. The resulting bundle of systems is topological in the sense described above, in full agreement with the gap labels. The edge spectrum can be resolved as function of $\omega \in \SM^1$ as in Fig.~\ref{Fig:EdgeExpVsTh}, by noticing that the coordinate of $\tau_a \omega$ on the circle is the angle, $a\theta$. The simulation in panel (a) shows that the edge spectrum splits into chiral bands, whose number equals the gap label $m$. Since the computation was performed on a finite rather than a halved system, the chiral bands appear always in pairs, one per edge.

\begin{figure}[b]
\center
  \includegraphics[width=\linewidth]{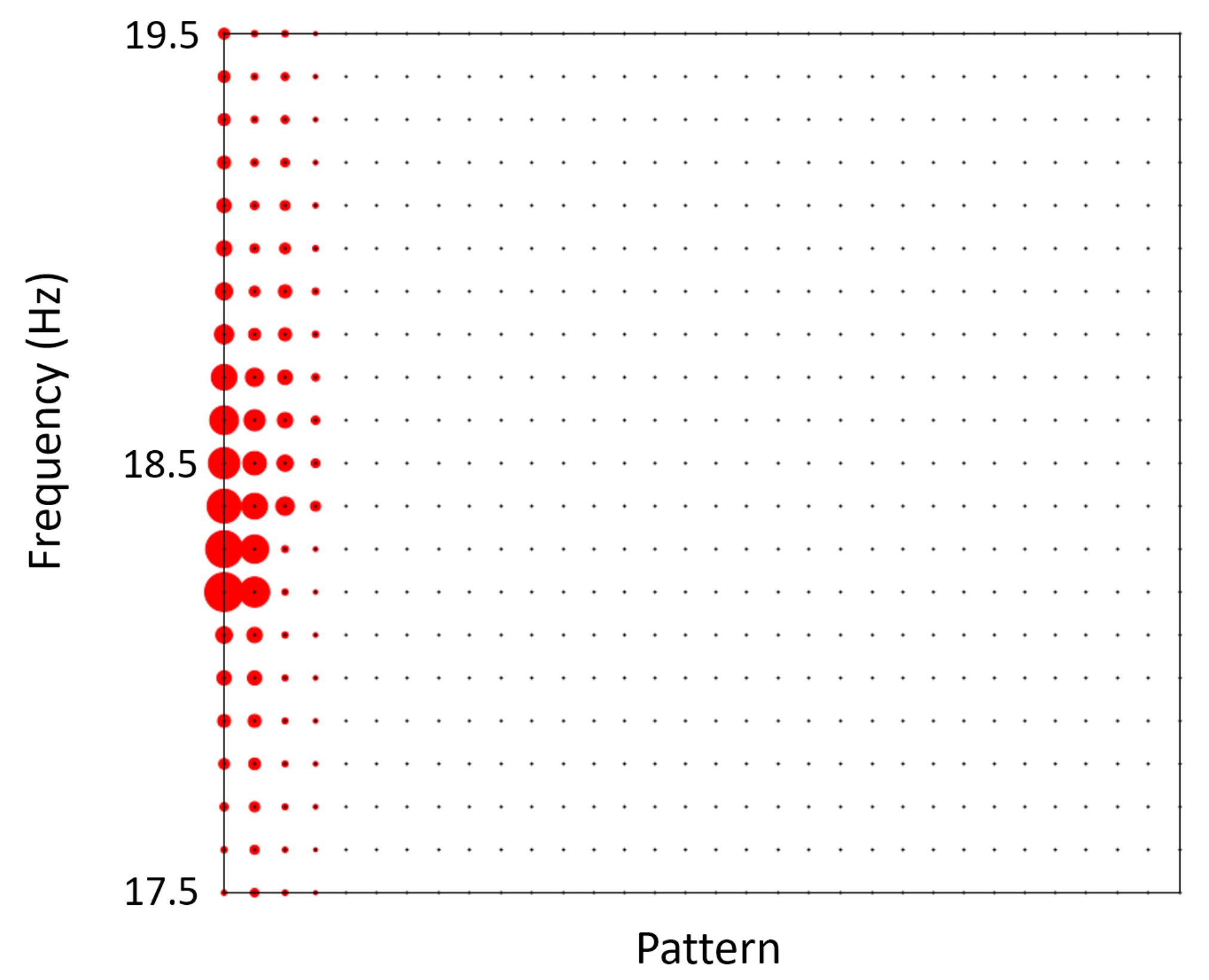}\\
  \caption{\small {\bf Spatial profile of a resonant mode.} The data reports the readings from four accelerometers placed on the first four spinners from the edge, as the frequency was swept over the last and most prominent bulk gap. The amplitudes of these readings are proportional with the size of the disks. For convenience, the full pattern of spinners is also shown.}
 \label{Fig:ResonantMode}
\end{figure}

\vspace{0.2cm}

The edge spectrum has been reproduced experimentally as reported in Fig.~\ref{Fig:EdgeExpVsTh}(b). In these experiments, the system shown in Fig.~\ref{Fig:BulkExpVsTh}(a) is actuated from the first spinner between 14 and 21~Hz in steps of 0.1~Hz. One spinner was then moved from the front to back of the chain, effectively implementing the translation $\tau_1\omega$, and the measurements were repeated. By cycling this whole process, one can shift the pattern 32 times and generate the experimental measure of the edge spectrum \eqref{Eq:StackedEdgeSpec}. Topological edge modes are detected at proper frequencies inside the bulk spectral gaps as the frequency is swept. They manifest as extremely strong and well-defined resonances, visible to the naked eye. A quantitative account of this phenomenon is reported in Fig.~\ref{Fig:EdgeResonance}, which displays the reading from an accelerometer placed on the arm of the second spinner from the edge. Fig.~\ref{Fig:ResonantMode} resolves the spatial profile of an edge resonant mode detected in the last and most prominent bulk gap. It confirms that the mode is extremely well localized near the edge.

\vspace{0.2cm}

\begin{figure*}[t]
\centering
\begin{subfigure}{0.3\textwidth}
\includemovie[poster=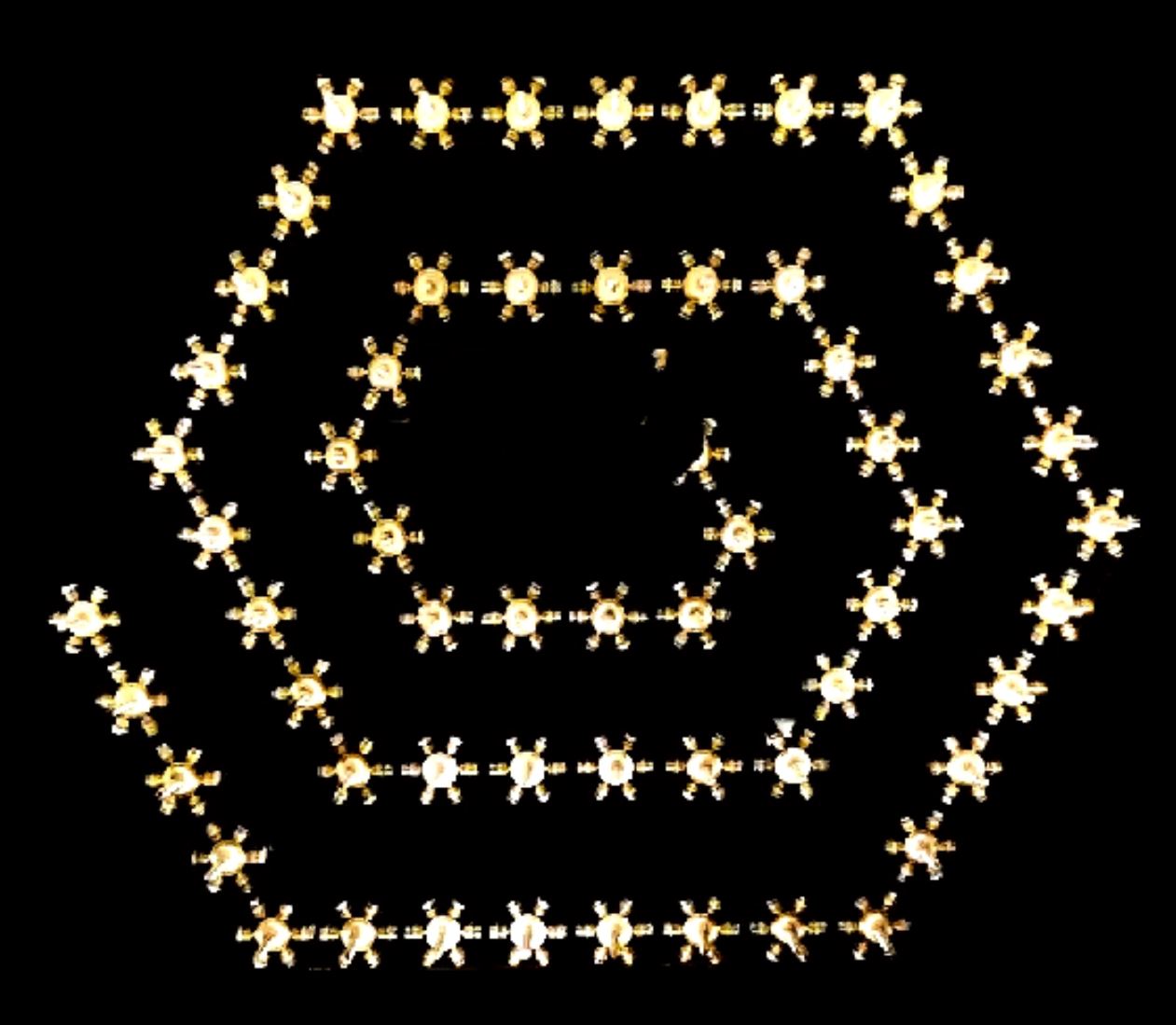,externalviewer,repeat=10,autoplay]{\linewidth}{\linewidth}{Bulk_Mode_11p2Hz_x0p25Speed.gif}
\end{subfigure}
\begin{subfigure}{0.3\textwidth}
\includemovie[poster=Poster1.jpg,externalviewer,repeat=10,autoplay]{\linewidth}{\linewidth}{Edge_Mode_19p5Hz_x0p25Speed.gif}
\end{subfigure}
\begin{subfigure}{0.3\textwidth}
\includemovie[poster=Poster1.jpg,externalviewer,repeat=10,autoplay]{\linewidth}{\linewidth}{Gap_Mode_21p5Hz_x0p25Speed.gif}
\end{subfigure}
\caption{\small Video recordings of the experimental system, illustrating a (left) bulk mode, (middle) topological edge mode, (right) off-resonance forced mode.}
\label{Fig:VideoRecordings}
\end{figure*}

Lastly, in Fig.~\ref{Fig:VideoRecordings} we report video recordings of the response of the experimental system when actuate from one end at different frequencies. In these experimental observations, the system has been enlarged to 64 spinners and the chain has been wrapped around in a spiral-like configuration. Since the excitations along the chain caries only angular and no linear momentum, there is no backscattering at the sharp corners. Another way of seeing this is by observing that the equations of motion remain unchanged. The first video recording exemplifies a bulk mode, where one can see a standing wave pattern over the entire structure, even when the system is actuated only at one end. In the second recording, the frequency has been tuned on the topological resonant mode occurring in the last spectral gap. To exemplify the difference between this resonant response and a trivial forced oscillation, we report a third recording where the driving frequency is off resonance.

\subsection{The algebra of half-space observables} For the half-space, the shift operator, which on the new Hilbert space will be called $\widehat S$, is no longer unitary, but instead: 
\begin{equation}
\widehat S \ \widehat S^\ast = I, \quad \widehat S^\ast \, \widehat S = I-P_0, \quad P_0= |0 \rangle \langle 0 |.
\end{equation} 
This suggests the definition of the half-space algebra, $\widehat \Aa$, as the algebra generated by $C_N(\Xi)$ and the operator $\hat u$ satisfying the same commutation relations \eqref{Eq:AlgCommRel}. However, $\hat u^\ast \hat u =1-\hat e$, with $\hat e$ a proper projection $\hat e^2=\hat e^\ast=\hat e \neq 1$. The half-space algebra accepts the following canonical representation:
\begin{equation}
C_N(\Xi) \ni f \rightarrow \hat \pi_\omega(f) = \sum_{n \geq 0} f(\tau_n \omega) \otimes |n \rangle \langle n |
\end{equation}
 and $\hat \pi_\omega(\hat u)= \widehat S$, which generates all half-space physical Hamiltonians \cite{ProdanSpringer2016}. In general, we have the isomorphism of algebras $\widehat \Aa \simeq \bigoplus\limits_{\omega \in \Xi} \hat \pi_\omega(\widehat \Aa)$ hence:
\begin{equation}
{\rm Spec}(\hat h) = \bigcup_{\omega \in \Xi} {\rm Spec}(H_\omega) = \overline{\bigcup_{a \in \ZM} {\rm Spec}(H_{\tau_a\omega})} = {\rm Spec}(\widehat H),
\end{equation}
with $H$ from \eqref{Eq:Stack}. In other words, the spectrum of $\hat h$ coincide with the spectrum of the stacked systems introduced and discussed above. This is important to keep in mind because the $K$-theoretic bulk-boundary principle contains a statement about the spectrum of $\hat h$, hence of the stacked systems, as already highlighted in Fig.~\ref{Fig:EdgeSpecTh}.

\vspace{0.2cm}

Inside $\widehat \Aa$, there is the ideal $\widetilde A$ made up of elements of the form $\hat a \, \hat e \, \hat b$ for some $\hat a, \hat b \in \widehat \Aa$. When represented on the physical space, such elements are localized near the boundary, hence $\widetilde \Aa$ is called the boundary algebra \cite{ProdanSpringer2016}. Given a bulk Hamiltonian, $h$, the half-space Hamiltonian with Dirichlet boundary condition, {\it i.e.} all couplings crossing the boundary set to zero, is generated by:
$$
\hat h_D=\sum_{q \geq 0}  h_{q} \hat u^q + \sum_{q < 0}  h_{q} (\hat u^\ast)^{|q|}. 
$$
By adding elements from $\widetilde \Aa$, $\hat h = \hat h_D + \tilde h$, the Dirichlet can be changed into any other boundary condition, hence the formalism is completely general. An important relation established in \cite{ProdanSpringer2016} is the isomorphisms between $\widetilde \Aa$ and $M_\infty(\CM) \otimes C(\Xi)$, which will play an important role for the bulk-boundary correspondence principle. That is because then $K_\ast(\widetilde \Aa) = K_\ast (C(\Xi))$, and since $\Xi \simeq \mathbb S^1$, $K_\ast(\widetilde \Aa) \simeq \mathbb Z$. In particular, $K_1(\widetilde\Aa)$ is generated by $[v]_1$, with $v$ introduced in \eqref{Eq:ExCommRel}.

\vspace{0.2cm}

\subsection{The engine of the bulk-boundary correspondence.} The following exact sequence between the algebras of physical observables is well established \cite{KRS-BRevMathPhys2002,ProdanSpringer2016}:
\begin{equation}\label{Eq-ExactSequence1}
\begin{diagram}
0 \rTo & \widetilde \Aa   & \rInto{i} &  \widehat \Aa  &\rOnto{ \mathrm{ev \ \ }}  &\Aa &\rTo 0,
\end{diagram}
\end{equation}
with ${\rm ev}(\hat u) = u$. This exact sequence sets in motion a six-term exact sequence at the $K$-theory level \cite{WOBook1993,BlaBook1998,RLLBook2000}:
\begin{equation}\label{Eq-SixTermSequence}
\begin{diagram}
& K_0(\widetilde \Aa) & \rTo{\ \ i_\ast \ \ } & K_0(\widehat{\Aa}) & \rTo{\ \ {\rm ev}_\ast \ \ } & K_0(\Aa) &\\
& \uTo{\rm Ind} & \  &  \ & \ & \dTo{\rm Exp} & \\
& K_1(\Aa)  & \lTo{{\rm ev}_\ast} & K_1(\widehat{\Aa}) & \lTo{i_\ast} & K_1(\widetilde \Aa) &
\end{diagram}
\end{equation}
The relevance of this diagram to the bulk-boundary principle program was established by the work of \cite{KRS-BRevMathPhys2002}. Examining the right side of the diagram, one can see the standard $K$-theory map ${\rm Exp}$ taking projections from the bulk algebra into unitaries from the boundary algebras. Given a gapped bulk Hamiltonian $(h,G)$ from $\Aa$, one defines a function $\phi : \RM \rightarrow \RM$ with a sharp, but continuous, variation in an $\epsilon$-interval around any point $G$ inside the gap, such that $\phi=0/1$ below/above that interval. Besides these requirements, $\phi$ is completely arbitrary. If $\hat h$ is any half-space Hamiltonian obtained from $h$, {\it i.e.} ${\rm ev}(\hat h) = h$, then \cite{KRS-BRevMathPhys2002,ProdanSpringer2016}:
\begin{equation}\label{Eq-ExpMap}
{\rm Exp}[p_G]_0 = [\tilde u_G]_1, \quad \tilde u_G=e^{2 \pi i \phi(\hat h)}.
\end{equation} 
As the notation suggests, $\tilde u_G$ is a unitary element of the boundary algebra, $\widetilde \Aa$, because the function $e^{2 \pi \epsilon}-1$ is non-zero only on the edge spectrum. Hence, $e^{2 \pi \hat h}-1$ is built only from boundary states. If the half-space Hamiltonian $\hat h$ is gapped, then we can take $G$ and the variation of the function $\phi$ inside this gap, in which case $\phi$ is either 0 or 1 on the spectrum of $\hat h$. By the rules of functional calculus, $\tilde u_G$ is the identity. If $\hat h$ is gapped, then $[p_G]_0$ is necessarily mapped into the trivial $K_1$-class $[1]_1$ by the ${\rm Exp}$ map. However, if the latter is not the case, then $\hat h$ cannot be gapped. The key conclusion is that, if ${\rm Exp}[p_G]_0 \neq [1]_1$, then  topological boundary spectrum emerges, filling the {\bf entire} bulk gap. This spectrum cannot be gapped by any boundary condition or adiabatic deformation of $h$.

\vspace{0.2cm}

To summarize, in order to establish a bulk-boundary correspondence principle, one needs to resolve the $K$-theories of both bulk and boundary algebras, as well as the action of ${\rm Exp}$ map on the generators of $K_0(\Aa)$. We should warn the reader that just the condition $[p_G]_0 \neq [0]_0$ is in general not enough, a counter example being the case of Fibonacci patterns where the whole $K_0$-group is mapped by the Exp map into the trivial $K_1$-class of the boundary \cite{KP2017}.

\subsection{Topological patterns, indeed.} From the bulk analysis, we know that $K_0(\Aa)$ is generated by $[1]_0$ and $[e_\theta]_0$ and that every bulk gap projection accepts a decomposition:
\begin{equation}
[p_G]_0 = n \, [1]_0 + m \, [e_\theta]_0, \quad m \neq 0.
\end{equation}
The action of the $\rm Exp$ map is also known \cite{ProdanSpringer2016} explicitly:
\begin{equation}
{\rm Exp}[1]_0 = [1]_1, \quad {\rm Exp}[e_\theta]_0 = [v]_1.
\end{equation}
One can see that any gap projection is mapped non-trivially: 
\begin{equation}\label{Eq:BB}
{\rm Exp}[p_G]_0 = m \, [v]_1 \neq [1]_1,
\end{equation} 
and, consequentially, topological edge spectrum fills every single spectral gap of a bulk Hamiltonian, regardless of its particular form. Furthermore, \eqref{Eq:BB} automatically implies that $\tilde u_G$ is homotopic to $v^m$, hence $m$ counts the winding of the eigenvalues of $\tilde u_G(\omega)$ as $\omega$ is varied along $\Xi \simeq \SM^1$. In turn, this tells that $m$ counts the number of chiral edge bands of $\hat h$, in agreement with the observations from Fig.~\ref{Fig:EdgeExpVsTh}.

\vspace{0.2cm}

In the cases when there are more internal degrees of freedom, $N>1$, the $IDS$ will take values in the interval $[0,N-1]$, hence there are $N-1$ possible instances where $m$ in \eqref{Eq:IDSRange} can be zero. As such, among the infinite number of bulk gaps there will be only $N-1$ gaps which are not topological, {\it i.e.}  will not be filled with edge spectrum when the system is halved. If there is enough control over the design of a meta-materials, of course, one should try to isolate just one degree of freedom per resonator but, if this is not possible, even the cases with large $N$'s will still display plenty of topological gaps.

\section{Discussion}

We conclude with proposals of how our findings can be incorporated in practical applications. First, let us recall that the topological system singled out by our work consists of bundles or stacking of translates of $H_\omega$. An important observation is that such bundles can be obtained by a simple and practical procedure. Indeed, by sequentially cutting equal pieces of length $L$ from a single bulk sample and bundling them together, one is effectively generating $\bigoplus_{n \in \NM} \widehat H_{\tau_{nL}\omega}$ (provided $L$ is large enough). Since $\tau_{nL}\omega$ is densely sampling the configuration space $\Xi$, the desired stacking has been achieved. 

\vspace{0.2cm} 

The bundle described above displays edge modes which cannot be removed by cutting, wear and tear, or by gentle bending of the strands. Additionally, the modes can be localized not only in space but also in frequency. By examining the spectral butterfly in Fig.~\ref{Fig:BulkSpecIDS}, one can see that, by varying $\theta$, one can align at least one bulk spectral gap at any desired frequency within the bulk range.

\vspace{0.2cm}

Let us end by noting that our conclusions are not bound to mechanical systems and they apply to any coupled resonators regardless of their nature. Hence, whenever spatial and frequency control over the excitation modes is desired, the proposed patterns can provide a convenient practical solution given the minimal tuning required.


\end{document}